\newcommand{\rep}[1]{{ #1}}

\documentclass[useAMS,usenatbib]{mn2e}
\usepackage{graphicx}
\usepackage{color}
\usepackage{ulem}
\title[WR\,63 : A multiple system (O+O)+WR~?]{WR\,63 : A multiple system (O+O)+WR\,?\thanks{Based on observations obtained at the Gemini Observatory, which is operated by the Association of Universities for Research in Astronomy, Inc., under a cooperative agreement with the NSF on behalf of the Gemini partnership: the National Science Foundation (United States), the National Research Council (Canada), CONICYT (Chile), the Australian Research  Council (Australia), Minist\'erio da Ci\^encia, Tecnologia e Inova\c c\~ao (Brazil) and Ministerio de Ciencia, Tecnolog\'ia e Innovaci\'on Productiva (Argentina).}}
\author[Chen{\'e}, Mahy, Gosset, et al.]{Andr{\'e}-Nicolas Chen{\'e}$^{1,2}$\thanks{E-mail:
andre-nicolas.chene@noirlab.edu}, Laurent Mahy$^{3}$\thanks{E-mail:
laurent.mahy@oma.be}, Eric Gosset$^{4}$, Nicole St-Louis$^{5}$, \and Karan Dsilva$^{6}$, and Rajeev Manick$^{7}$\\
$^{1}$ Gemini Observatory/NSF’s NOIRLab, 670 N. A‘ohoku Place, Hilo, Hawai‘i, 96720, USA\\
$^{2}$ Visiting astronomer at the Universit{\'e} de Montr{\'e}al, Complexe des Sciences, Montr{\'e}al, QC H2V 0B3, Canada\\
$^{3}$ Royal Observatory of Belgium, Avenue Circulaire/Ringlaan 3, B-1180 Brussels, Belgium\\
$^{4}$ Space sciences, Technologies and Astrophysics Research (STAR) Institute, Universit{\'e} de Li{\`e}ge, All{\'e}e du 6 Ao{\^u}t 19c (B5c), B-4000, Li{\`e}ge, Belgium\\
$^{5}$ Centre de Recherche en Astrophysique du Qu{\'e}bec, D{\'e}partement de physique, Universit{\'e} de Montr{\'e}al, Complexe des Sciences, Montr{\'e}al, QC H2V 0B3, Canada\\
$^{6}$ Institute of Astronomy, KU Leuven, Celestijnenlaan 200 D, B-3001 Leuven, Belgium\\
$^{7}$ Institut de Plan{\'e}tologie et d’Astrophysique de Grenoble, Univ. Grenoble Alpes, 38000 Grenoble, France\\}
\begin{document}

\date{Accepted 2022 XXX YY. Received 2022 October 01; in original form 2022 October 01}

\pagerange{\pageref{firstpage}--\pageref{lastpage}} \pubyear{2022}

\maketitle

\label{firstpage}

\begin{abstract}
The spectrum of the Wolf-Rayet (WR) star WR\,63 contains spectral lines of two different O stars that show regular radial velocity (RV) variations with amplitudes of $\sim$160 and $\sim$225 km\,s$^{-1}$ on a $\sim$4.03\,d period. The light-curve shows two narrow eclipses that are 0.2\,mag deep on the same period as the RV changes. On the other hand, our data show no significant RV variations for the WR spectral lines. Those findings are compatible with WR\,63 being a triple system composed of two non-interacting late O stars orbiting a WR star on a period larger than 1000 days.
The amplitude of the WR spectral line-profile variability reaches 7--8\% of the line intensity and seems related to a 0.04\,mag periodic photometric variation. Large wind density structures are a possible origin of this variability, but our data are not sufficient to verify this. 
Our analysis shows that, should the three stars be bound, they would be coeval with an age of about 5.9$\pm$1.4\,Myrs. The distance to the O stars is estimated to be $3.4\,\pm\,0.5$\,kpc. Their dynamical masses are $14.3 \pm 0.1$ and $10.3 \pm 0.1$\,M$_\odot$. Using rotating, single star evolutionary tracks, we estimate their initial masses to be $18 \pm 2$ and $16 \pm 2$\,M$_\odot$ for the primary and the secondary, respectively. Regular spectral monitoring is required in the future to detect RV variations of the WR star that would prove that it is gravitationally bound to the close O$+$OB system and to determine its mass.
\end{abstract}


\section{Introduction}


\rep{Massive stars are one of the most important cosmic engines in the galaxies. During their evolution, they radiatively and mechanically influence their surroundings with their stellar winds, strong UV ionizing fluxes and their deaths as supernovae. The WR stage represents one of the most advanced phase in massive star evolution. In this phase, massive stars are characterized by strong emission feature spectra from highly ionized species \citep[see][for a detailed review]{crowther07}. They are classified in three different categories depending on their degree of stripping starting from the nitrogen-rich WR stars (WN), the carbon-rich WR stars (WC), and finally, the oxygen-rich WR stars (WO). These objects might be formed through single-star evolution (classified as main-sequence WR stars, or classical WRs which have evolved off the main-sequence through self stripping; \citealt{conti76,smith14}) or through binary interaction (through mass transfer or merging in binary or multiple systems; \citealt{paczyncki67, shenar20}).

There are currently more than 660 Wolf-Rayet (WR) stars known in our Galaxy \citep[][]{vH01,rosslowe15}\footnote{http://pacrowther.staff.shef.ac.uk/WRcat/index.php}, but only a few of them were revealed to be truly core-eclipsing binaries\rep{, i.e., systems where the eclipses occur when the projected stellar disks on the sky occult one another}}. Those are WR\,20a \citep{Ra04,Bo04}, WR\,22 \citep{Ba89,Go91,Ra96}, WR43a \citep[a.k.a. NGC\,3606-A1;][]{Mo04,Sc08}, WR\,139 \citep{Ga41,Ma94}, WR\,151 \citep{Hi48,Le93,Hu09} and WR\,155 \citep{Ga44,Ma95}. And, in the Large Magellanic Cloud, there is R144 \citep{shenar21}. For all these systems, the basic stellar parameters could be determined with high accuracy, and the orbit well defined. Their double eclipsing nature (except for WR\,22 which shows only one eclipse because of the high eccentricity of its orbit) offers the opportunity to directly measure the masses of their components by simple, least model-dependent Keplerian orbit.

WR\,20a and 43a are WN+WN systems and the two components are hydrogen-rich WR stars with masses over 80\,M$_\odot$. WR\,22 is a WR+O system, and the WR component is also an hydrogen-rich star with a mass in the range 56 to 72\,M$_\odot$ \citep{Ra96,Sc99}. The three other systems are WR+O systems, and the WR component is a classical, evolved WR star which has a mass around 15--20\,M$_\odot$. Such reliable determinations of WR star masses are of very high value for constraining and validating \rep{evolutionary} models. \rep{Indeed, binary evolution \citep{vanbeveren98, Sa12,moe17}, (rotational) mixing \citep{Me03}, and mass loss \citep{smith14,vink21} are key ingredients that still suffer from a good knowledge to understand massive star evolution. Therefore, while the Geneva and Bonn models (\citealt{ekstrom12} and \citealt{brott11}, respectively) for single-star evolution as well as the Binary Population and Spectral Synthesis (BPASS) models \citep[][]{St18} for binary evolution are commonly used, they are still limited by many simplifying assumptions \citep[see][for further details]{martins13}.
Those can be better put to test with well constrained systems. Another utility of core-eclipsing binaries is that they allow for stellar parameters derived from atmospheric models to be accurately associated with stellar masses.}


\rep{The present paper focuses on WR\,63 a WN7o star that} has been suspected of having an O5V or B0III companion based on its diluted emission lines \citep[][and related catalogs]{vH01}. Also, \citet{Ra20} were unable to determine the {\it{Gaia}} distance to the star because \rep{it is absent from the catalog, probably due to} the presence of that putative companion. Here, we present the discovery of two O-star spectra blended with the WR spectrum, indicating the possible multiple nature of WR\,63. It is not analogous to the binaries presented above, as the WR star is not eclipsing, but the two O stars are. In this work, we present the data as if the three stars were bound as a starting hypothesis. From the data in hand, we can analyse the photometric variations and radial velocity (RV) changes caused by the O stars and determine some of their fundamental parameters. We also present our analysis of the WR star line-profile variability and RV changes, and how future observations could lead to determine its mass.

In Section\,\ref{Observations}, we present our new photometric and spectroscopic data. In Section\,\ref{Analysis}, we describe our methods to \rep{search for periods, measure RV and determine line-profile variations (LPV)}. In Sections\,\ref{binary_param} and \ref{Evolution} we present the details of the O+O inner binary and the evolution of the system, respectively. Finally, we present our conclusions in Section\,\ref{Conclusions}.

\section[]{Observations}\label{Observations}
 
\subsection[]{Photometric observations}
As part of an intensive multi-band photometric monitoring campaign of WR stars with the Swope 1.0m telescope (Las Campanas), we have observed WR\,63 between May 12th and 31st 2012. Only 3 nights were lost due to bad weather over the entire run. Series of 10 CCD frames were obtained in each $B$, $V$ and $I$-bands 2 to 3 times per night, using the 1200$^2$ pix SITe\#3 detector. Each series was separated in time by at least 2.5h. We used the {\sc idl} functions {\it{find}} to detect the sources in the field and {\it{aper}} to perform the photometric measurements. The aperture was selected to be 3.5 times the full width at half maximum (FWHM) of the mean point-spread function (PSF) of the sources (hence, is dependent on the seeing conditions), and the sky annulus was taken from 3.5 to 7 times the FWHM of the mean PSF of the sources. In the 8.7\,arcsec$^2$ field-of-view of the SITe\#3 detector, we have detected 125 stars in $B$, 256 stars in $V$ and 806 stars in $I$. To perform relative photometry and correct for the effects of variable weather conditions, we select the frame taken under the best conditions (greatest number of sources detected, and a small value of the average PSF FWHM), and we compare the magnitude measured in that frame with the magnitudes measured in all the frames. The difference measured is assumed to be equal to the shift of the zero point in magnitude that we must apply to bring all the frames to the same level.

The Transiting Exoplanet Survey Satellite (TESS, \citealt{ricker14}) light-curve was taken in short cadence mode in Sector 38. We retrieved the light-curves, from the Mikulski Archive for Space Telescopes (MAST3) archive, in its pre-conditioned form (PDCSAP, Pre-search Data Conditioning Simple Aperture Photometry). We detrended it using a linear fit to remove any long-term trends. The light-curve covers heliocentric Julian dates (HJD) from $\mathrm{HJD} = 2\,459\,334$ to $\mathrm{HJD} = 2\,459\,361$. The pseudo-Nyquist frequency is $\nu_{\rm Ny} = 357.143$~d$^{-1}$ (leading to a step of 241.92 s). The word “pseudo” is used to point out the fact that the aliasing is not pure.

\subsection[]{Spectroscopic observations}
We collected high-resolution spectra with the Fibre-fed Extended Range Optical Spectrograph \citep[FEROS,][]{kaufer97, kaufer99} mounted on the MPG/ESO 2.2m telescope at La Silla (Chile). FEROS provides a resolving power of $R = 48,000$ and covers the entire optical range from 3800 to 9200\,\AA. The data (Program ID: 089.D-0730, PI: Mahy) were obtained from 2012 May 1st to 2012 May 7th. Their reduction was done following the procedure described in \citet{mahy10,mahy17}. 

We collected three spectra with the UV and Visible Echelle Spectrograph \citep[UVES,][]{dekker00} mounted on the ESO/VLT UT2 (Paranal, Chile). These three spectra were collected in 2017 Apr. 22nd (Program ID: 099.D-0314, PI: Mahy), 2018 May 29th (Program ID: 0101.D-0273, PI: Mahy), and 2019 May 22nd (Program ID: 0103.D-0414, PI: Mahy), with CD\#3-520nm, CD\#3-520nm, and CD\#3-564nm setups, respectively. These setups cover the $4143-6214$\,\AA, and $4583-6686$\,\AA\ wavelength ranges. The resolution of UVES is $R \sim 85,000$. The UVES data were reduced using the ESO pipeline. 

Optical spectroscopic time series were also obtained using the High Resolution Spectrograph (HRS) on SALT \citep{bramall10,bramall12,crause14} under program ID 2021-1-SCI-013 (PI: Manick). The data were taken in low-resolution mode ($R \sim 16,000$ ) and cover the $3735-8779$\,\AA\ wavelength domain. The data were reduced with the \textsc{midas} pipeline \citep{kniazev2016} based on the {\it {\'e}chelle} \citep{ballester92} and {\it feros} \citep{stahl99} packages. We applied heliocentric corrections to the data.

Finally, lower resolution spectra were obtained at Gemini using the GMOS-South spectrograph \citep{Ho04} under programs GS-2015A-Q-88, GS-2014A-Q-94, GS-2019B-Q-404 and GS-2020A-Q-410 (PI: Chen\'e). The majority of the spectra were obtained with the B600 grating, offering a resolution power of $R\sim1700$ over a spectral range of $3855-6980$\,\AA, while four spectra were obtained with the R831 grating covering $3880-5050$\,\AA\ with $R\sim3800$. The data were reduced using the Gemini package in \textsc{iraf}\footnote{{\sc iraf} was distributed by the National Optical Astronomy Observatory, which was managed by the Association of Universities for Research in Astronomy (AURA) under a cooperative agreement with the National Science Foundation.}.

\section[]{Analysis}\label{Analysis}
\rep{A first inspection of the Swope light curve revealed strong eclipses. The spectroscopic monitoring uncovered absorption lines related not just to one, but to two O stars. The following describes in detail our determination of the orbital motions.}

\subsection{Search for photometric periods}\label{photo}

\begin{figure}
\includegraphics[width=8.4cm]{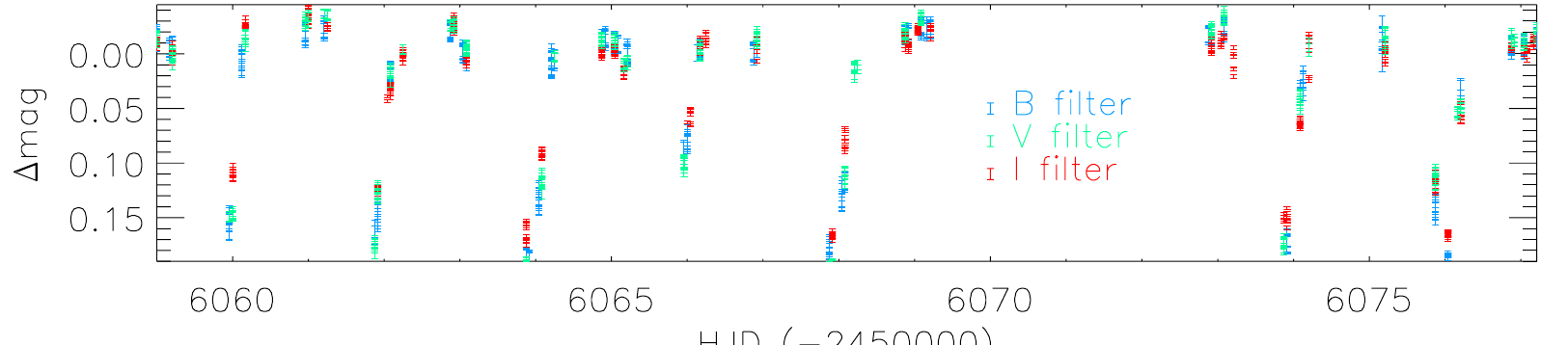}
\includegraphics[width=8.4cm]{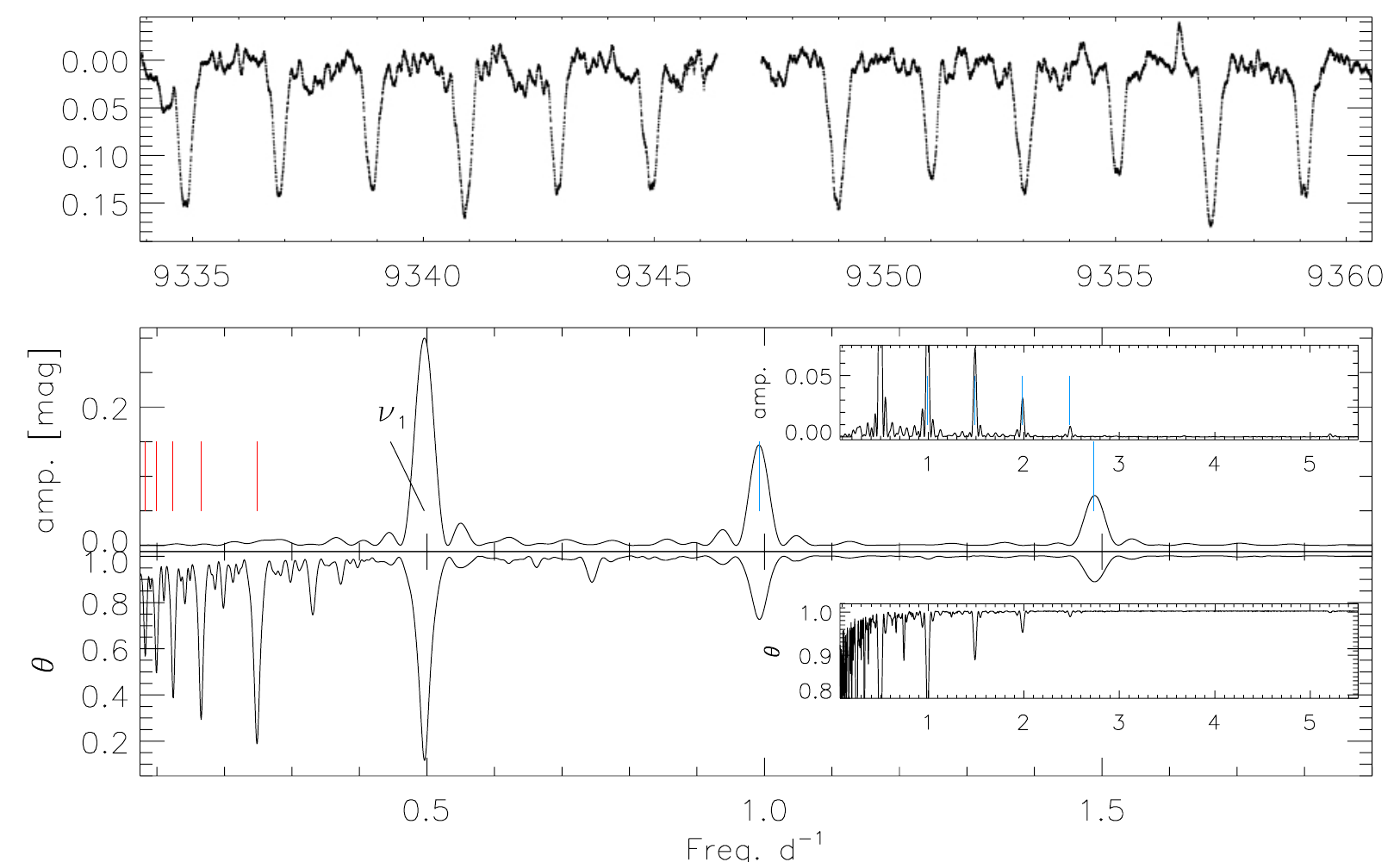}
  \caption{Swope (top panel) and TESS (middle panel) light-curves of WR~63. Bottom: The semi-amplitude Fourier periodogram of the TESS light-curve and the $\theta$ spectrum from the PDM method. The highest peak is marked as $\nu_1$. The associated harmonics are marked with blue lines (for $N\times\nu_1$) and the PDM subharmonics with red lines (for $\nu_1/M$).} \label{periodo}
\end{figure}

We ran a Fourier analysis based on the Heck-Manfroid-Mersch technique \citep[][revised by \citealt{gosset01}]{heck85} on the Swope and the TESS light-curves. We selected the highest peaks based on the criterion presented by \citet[][their equation 2]{mahy11}. The results from the Swope light-curve is similar, but of course much noisier than the results from the TESS light-curve. This iterative criterion provides us with about 60 formally significant frequencies. This large number is due to the large content of deterministic variability in the light-curve. The semi-amplitude spectrum is dominated by a set of frequencies distributed according to a regular pattern including the fundamental frequency at about 0.498(1)~d$^{-1}$ (denoted as $\nu_1$ in Fig.\,\ref{periodo}) and its harmonics at $N\times\nu_1$ with $N=2,3,4$ and 5. The presence of the harmonics is easily explained by the rather narrow shape of the eclipses. We also detect a significant frequency at 0.251(1)~d$^{-1}$ but it is unclear whether that frequency is connected to the fundamental frequency mentioned above or to an independent signal since the span of time of the TESS data is too short. The presence of this frequency most probably indicates that the eclipses present different depths in alternance.  

\begin{figure}
\includegraphics[width=8.4cm]{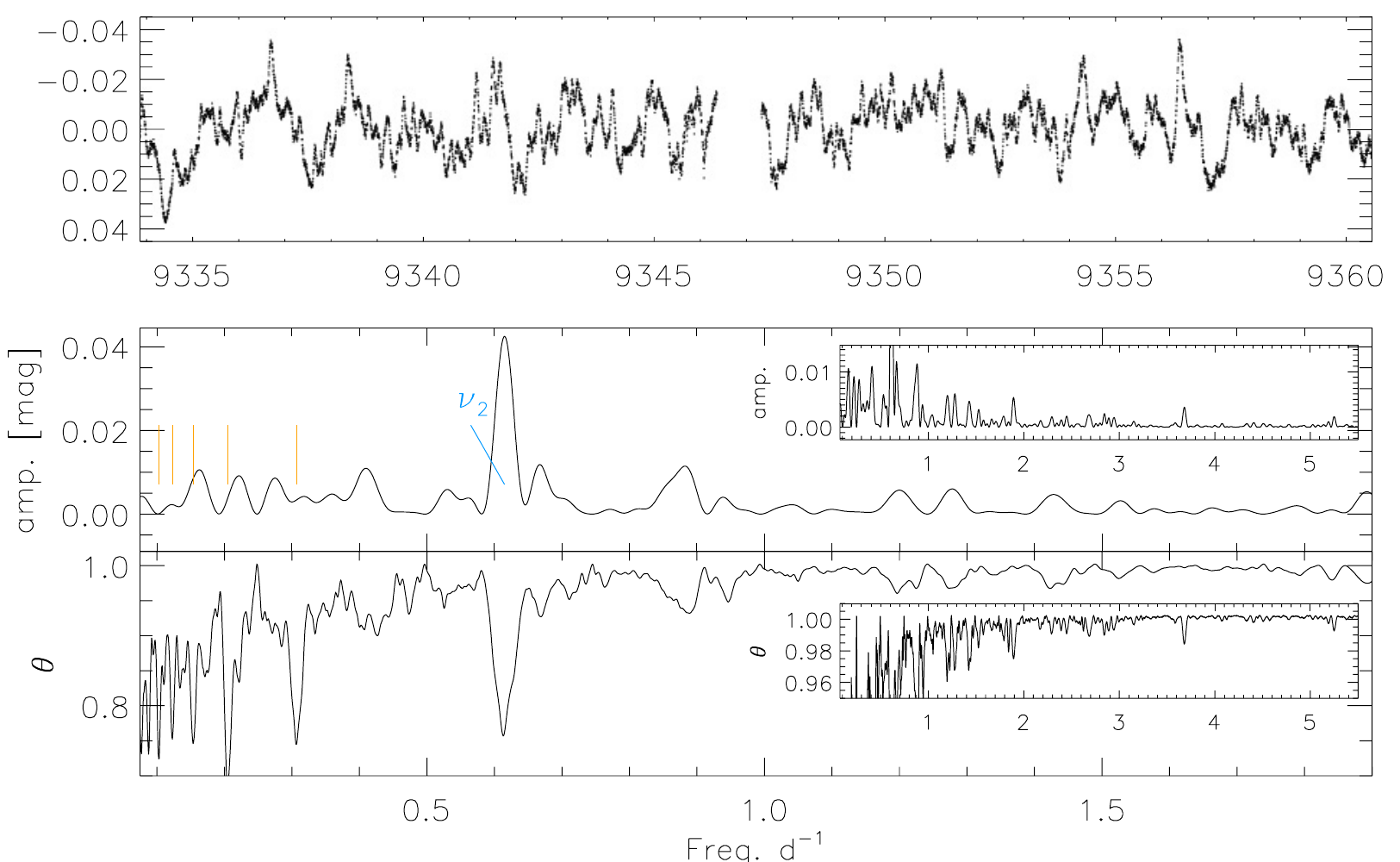}
  \caption{Same as Fig.\,\ref{periodo}, but for the TESS light-curve after the eclipses have been subtracted.}
  \label{resTESS}
\end{figure}

We also performed a period search using a phase-dispersion minimization (PDM) algorithm \citep{St78} which is well-suited to cases in which the signal is highly non-sinusoidal, such as core eclipses. The method consists in folding the data points in phase using different trial periods and dividing the resulting curves in a predetermined number of phase bins. For each trial period, we can define $S_j^2$, the sample variance around the mean for bin $j$, $S^2$, the cumulated variance over all $S_j^2$ and finally $\theta = S^2/\sigma^2$, where $\sigma^2$ is the variance of all data points. In addition to the main period $\nu_1$ and its harmonics, it also displays more peaks at $\nu_1/M$ with $M=2,3,4$ and 5 (see Fig.\,\ref{periodo}). These additional peaks are subharmonics and are a well-known artefact of the phase-folding minimization methods. These artefacts can be easily recognized thanks to the widths of the peaks that are narrower (width behaving as $1/M$).

\rep{The fundamental period is $P_{\rm orb} = 4.0275 \pm 0.0008$~d, and it corresponds to $\nu_1/2$. It also agrees with the RV motion of the He\,{\sc i} lines in absorption, which will be discussed shortly in Section\,\ref{RV}}. 

\begin{figure}
\includegraphics[width=4.15cm]{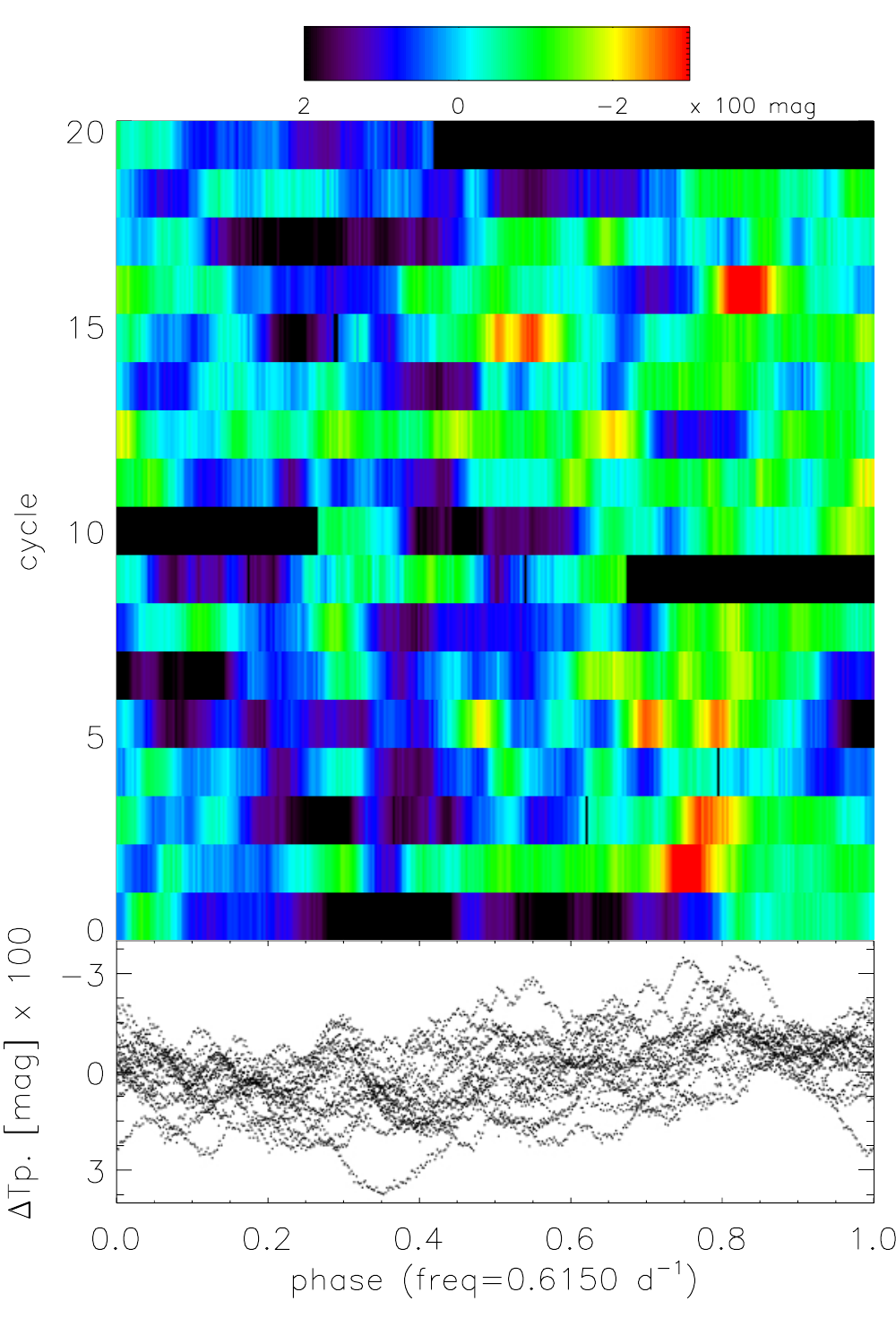}
\includegraphics[width=4.15cm]{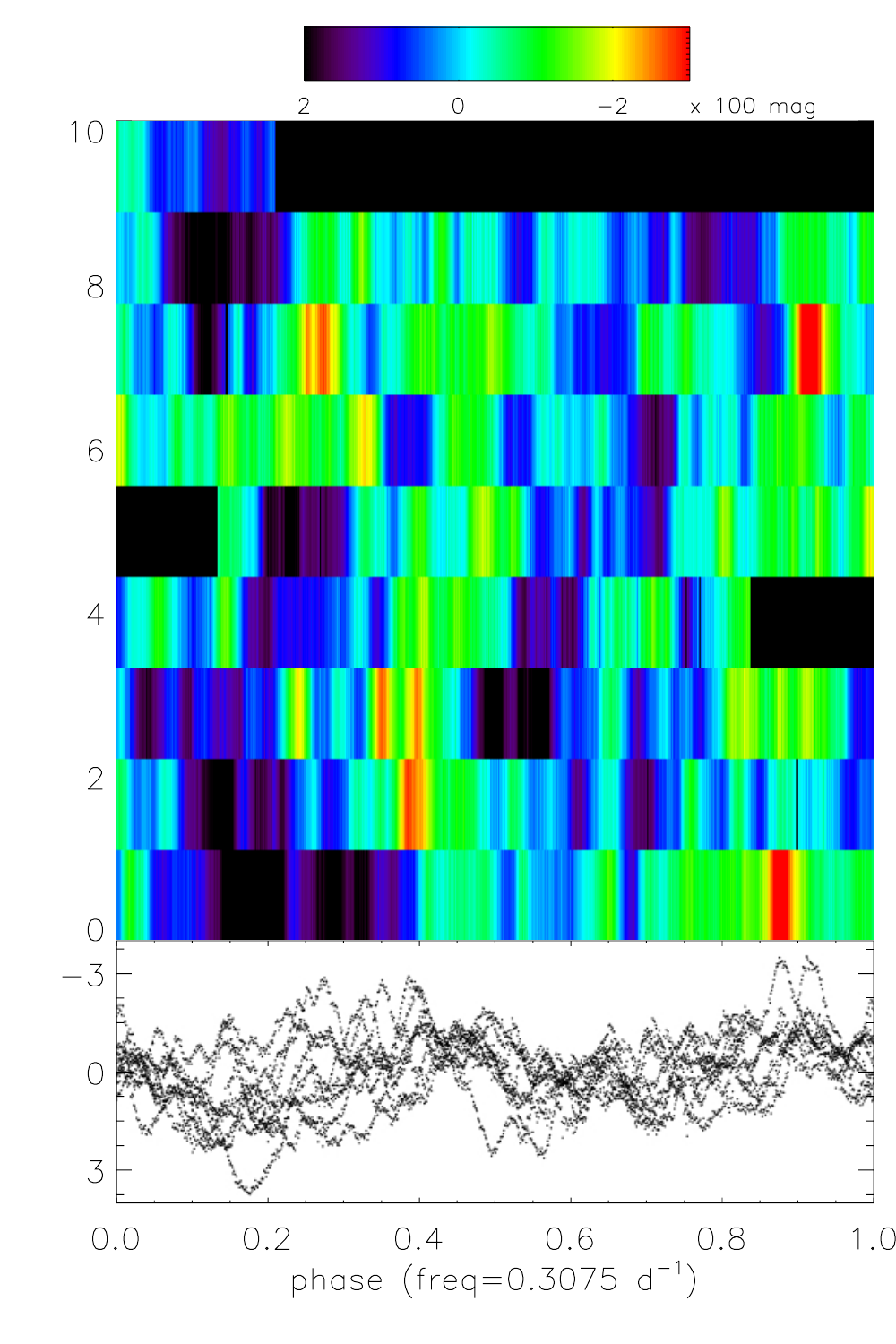}
  \caption{Top: TESS light-curve after subtraction of the eclipses, folded into the frequencies 0.6150~d$^{-1}$ (left) and 0.3075~d$^{-1}$ (right). Each cycle is shown in succession and the residuals are coded in color. Bottom: Same thing, but all the cycle overlaid.}
  \label{lcfold}
\end{figure}

After removing the signal of the eclipses from the TESS light-curve, we found a new peak at 0.615(1)~d$^{-1}$ (denoted as $\nu_2$ in Fig.\,\ref{resTESS}) and related peaks at $\nu_2/M$ with $M=2,3,4$ and 5. In Fig.\,\ref{lcfold}, we present the light-curve folded with the frequencies $\nu_2$ and $\nu_2$/2. It is very subjective to determine if there is any coherent features that repeat at every cycle and exploring more harmonics would leave a very little number of cycles to work with. \rep{Possible origins of these variations will be covered in Section\,\ref{lpv}}. 

Once the frequencies $\nu_1$ and $\nu_2$ are removed, we are left with a barely significant peak at 5.26(2)~d$^{-1}$, corresponding to a period of 0.192~d. This corresponds to the time span of a single smaller cusp, such as those pervading the entire TESS light-curve. 

\subsection{Radial velocity measurements}\label{RV}
The spectrum of the WR\,63 system is dominated by the WR star. Only a few absorption lines from the O companions can be found on top of strong emission lines. Also, large amplitude LPV (see Section\,\ref{lpv}) hinder the detection and the monitoring of strong photospheric lines expected in O stars, such as the lines of the Balmer and Pickering series. Nevertheless, He\,{\sc i} lines such as 4388\,\AA, 4922\,\AA\, and 5016\,\AA\, could be isolated fairly easily and were used to determine the RVs of the O stars. 

First, to increase the signal-to-noise ratio ($S/N$), which tends to be lower in the blue end of the spectra compared to redder wavelengths, we combined the three above listed He\,{\sc i} lines into a mean profile. The average was weighted by the $S/N$ measured in the continuum near the respective lines and was performed in velocity space. Second, we selected the spectra with the largest separation between the lines of both O stars, in order to fit their profiles using Gaussian functions. Those were then used as a reference profile for each of the O stars. Finally, for each spectrum, we fit the two components simultaneously by a simple minimisation of $\chi^2$. The uncertainties are determined by the interval in velocity that provides a $\chi^2$ near the minimum and within a range corresponding to the $S/N$ of the spectrum. 

\begin{figure}
\includegraphics[width=8.4cm]{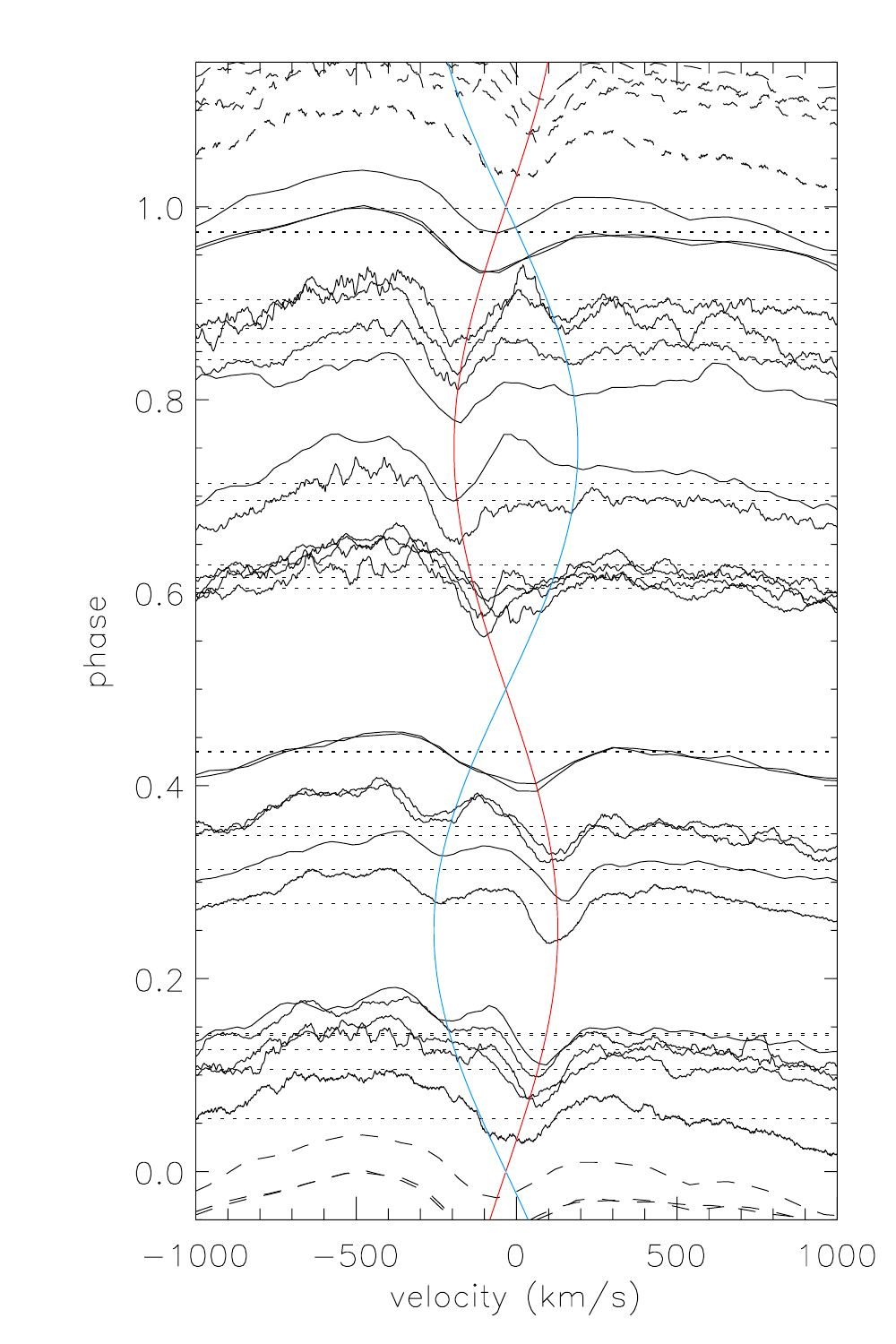}
  \caption{Mean profile of the He\,{\sc i} 4388\,\AA, 4922\,\AA\, and 5016\,\AA\ lines. Their position on the y-axis is determined by the orbital phase, based on the period determined in Section\,\ref{photo}. The orbital solution of Section\,\ref{binary_param} is plotted.}\label{HeI}
\end{figure}

\rep{In Fig.\,\ref{HeI}, the combined He\,{\sc i} line profiles are organized based on the \rep{photometric} period presented in Section\,\ref{photo}. The RV curves of the two O stars are shown to guide the reader's eye to the centroids of the absorption lines. The two O-star RV semi-amplitudes are $160\pm15$\,km\,s$^{-1}$and $225\pm15$\,km\,s$^{-1}$. The sinusoidal shape of the RV curve, combined with the light-curve showing eclipses being of identical widths and separated by exactly half the orbit period, suggest that the orbit of the O$+$O system is circular.

Determining the RV of the WR star requires another strategy.} There are no true atmospheric lines in the WR spectrum and the majority of the emission lines are affected by the strong LPV \rep{(see Section\,\ref{lpv})}. Nevertheless, it is possible to minimise the effect of LPV by selecting a spectral line of high ionisation level. Indeed, since those lines are formed in a region closer to the surface of the star, it is expected that the impact of wind density structures at the origin of the LPV is significantly smaller than for lines formed further out in the wind. We therefore selected the N\,{\sc iv}\,4058 for this purpose, \rep{as it is usually formed twice as close to the base of the WR wind than any other optical lines presented here \citep{Hi88}.} We used a cross-correlation method, bisector measurement and a simple fit of a Gaussian function to the line to determine RVs. The cross-correlation was done through three iterations; the first one using the mean profile of all the spectra, the next two ones using the mean again, but after each individual spectrum was corrected by the previous measurement of the velocity shift.

All the methods give the same RV curve for the WR star, within the uncertainties. Unfortunately, the GMOS-South spectra using the B600 grating have an RV precision that is of the same order of magnitude as the RV variations of the WR star, and were therefore removed from the series.

\begin{figure}
\includegraphics[width=8.4cm]{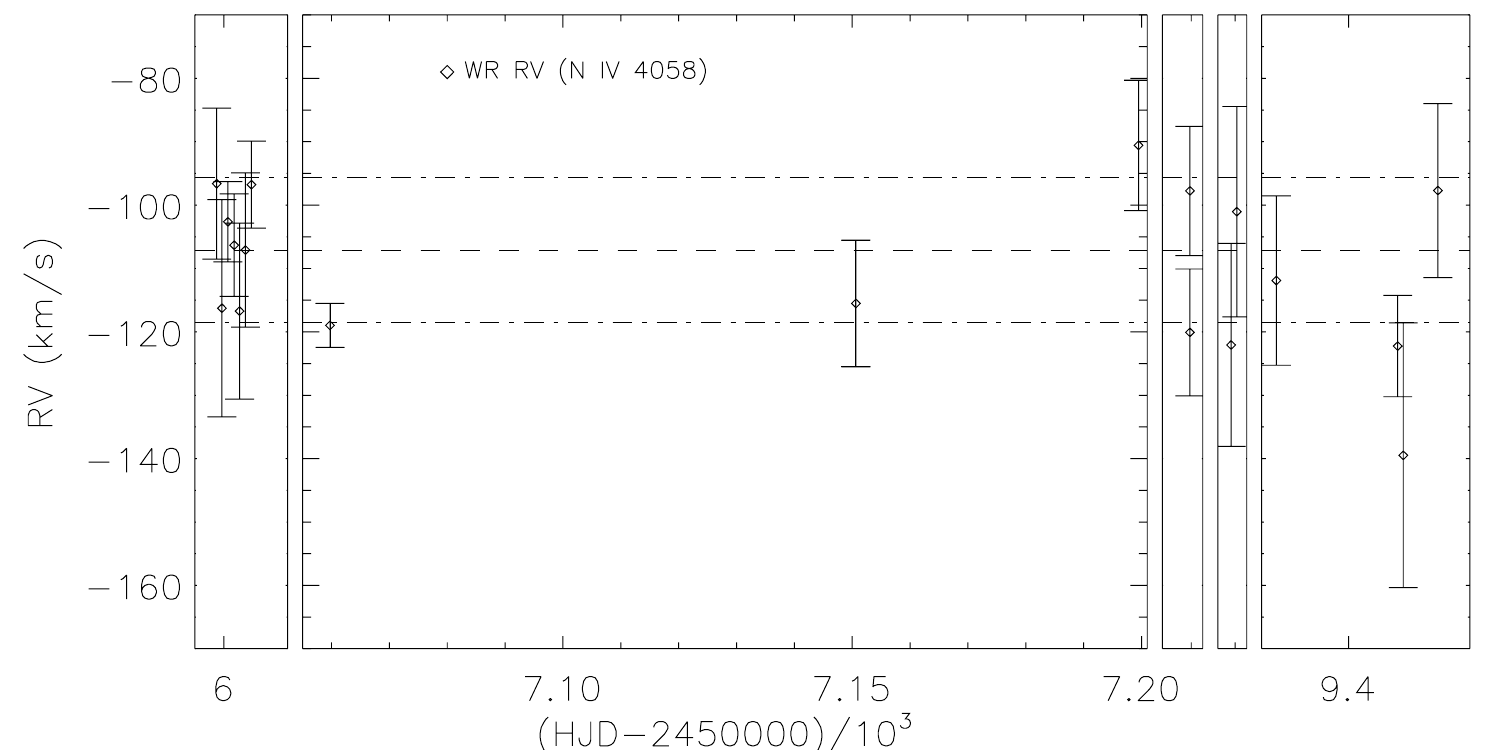}
  \caption{RV of the WR star (black) as a function of time. The average RV of the WR and its uncertainties are drawn with a dashed and dash-dotted lines, respectively.}\label{WRRV}
\end{figure}

Basically, through all of our spectroscopic campaigns, the WR star was static at an average RV near $-106\pm23$\,km\,s$^{-1}$ (see Fig.\,\ref{WRRV}). This is consistent with a triple system formed of two close O stars orbiting the WR star with a high eccentricity and/or a long period. Further spectra would therefore be necessary to hunt for a potential periastron passage. Interestingly, the RV values of the center of mass of the two O stars do not follow exactly the WR RV values. This will be discussed later in this publication.

\subsection{WR wind variability}\label{lpv}
The wind of the WR star in the WR\,63 system is known to be highly variable. Indeed, \citet{Ch11b} have shown that its LPV can reach an amplitude as high as 10\% of the spectral line intensity. These variations are expected to be caused by density wind structures at small and/or large scale. Small-scale wind structures are expected in virtually all WR stars due to wind clumping \citep{Mo88}. On the other hand, periodic, large amplitude LPV have been found in some WR stars such as WR\,1, 6, 110 and 134 \citep[][respectively]{Ch10,Mo97,Ch11a,Mc94}. These are often thought to be due to the presence of corotating interaction regions (CIRs) in their winds, and that possibility is going  to be our working assumption in this study. 

\begin{figure}
\includegraphics[width=8.4cm]{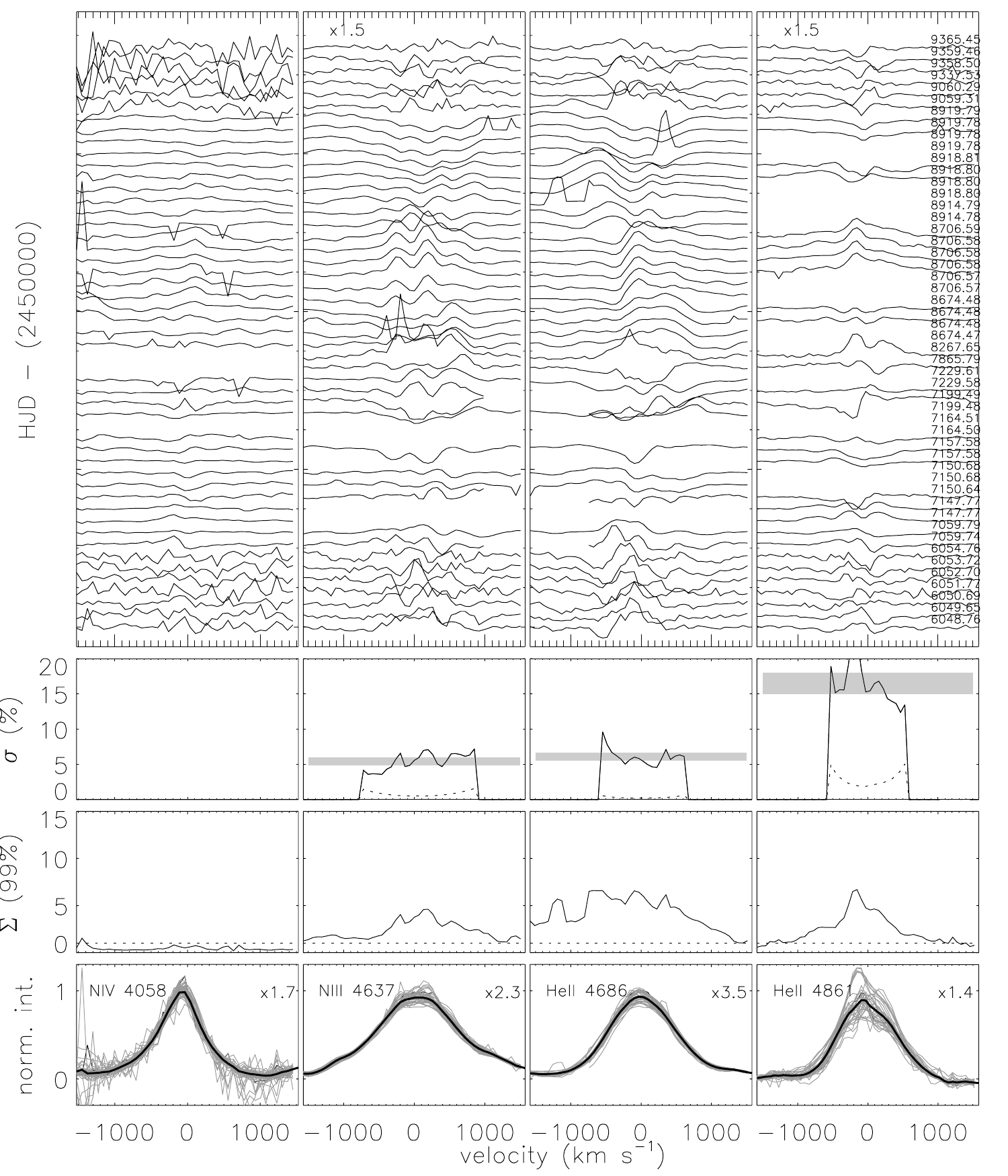}
\includegraphics[width=8.4cm]{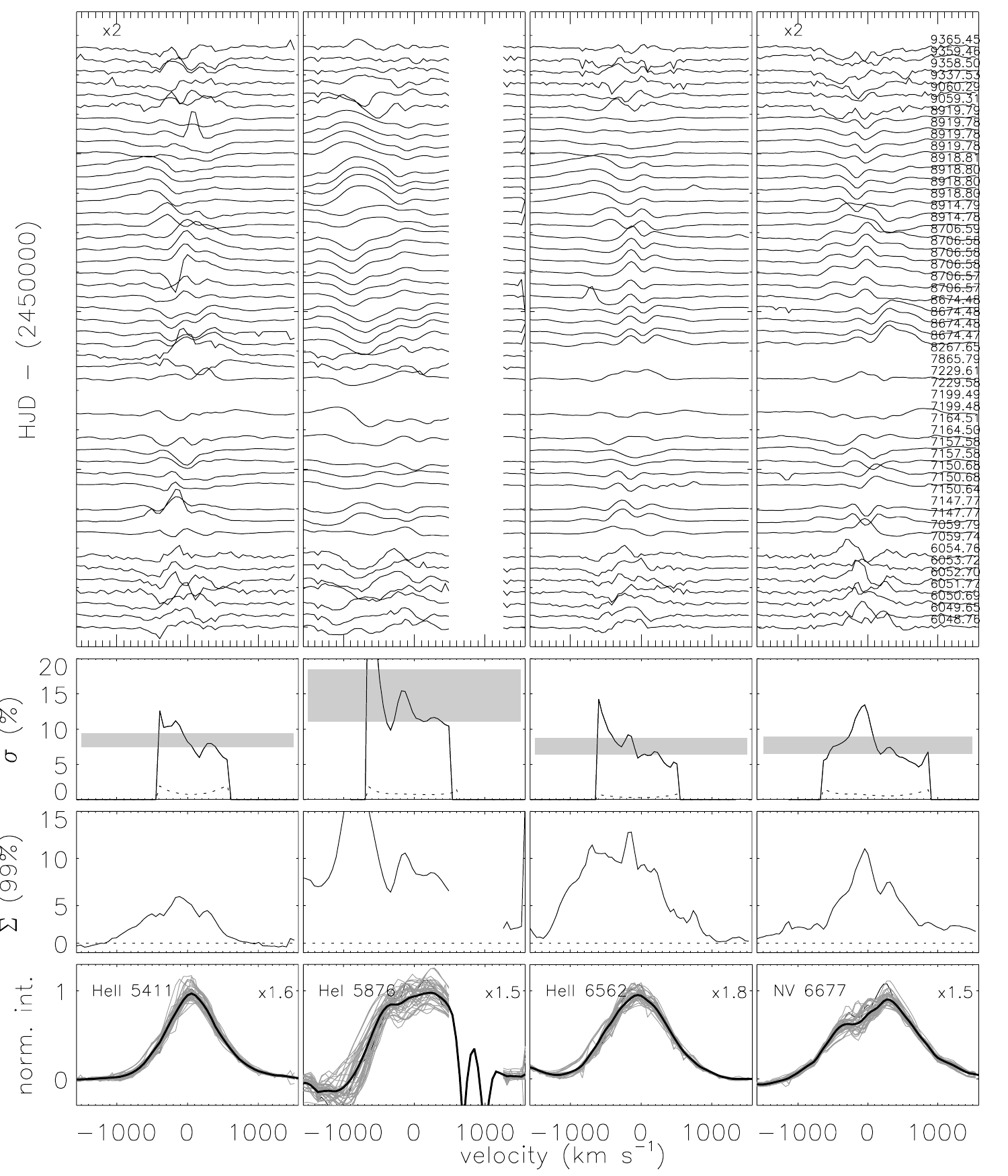}
  \caption{Variability in 8 emission lines from the WR wind. The bottom section shows the average profile of the lines with every individual profile of all the observing runs superposed in pale gray. The top section shows the residuals of each individual spectra with the average profile. The HJD is indicated for each residual on the right of the plot. In the middle two sections the $\Sigma (99\%)$ and the $\sigma$ (\%) spectra are presented; see text for more details.}
\label{fig:lpv}
\end{figure}

We analysed our spectra to highlight LPV in WR\,63. Since we know that the emission lines are diluted by the continuum of the O stars, we adjusted the spectra to take into account the variable continuum due to the eclipses (presented in Section\,\ref{binary_param}). In Fig.\,\ref{fig:lpv}, we present the residuals of each spectrum obtained by subtracting the mean profile (top) from individual spectrum (bottom). In the middle panels we display the Temporal Variance Spectrum (TVS) as defined by \citet{Fu96} and expressed as $\Sigma (99\%)$ ($=\sqrt{TVS/[\sigma_0^2 \chi_{N-1}^2(99\%)]}\,$, where $\sigma_0$ is the reciprocal of the rms of the noise level in the continuum in a time-series of $N$ spectra) and $\sigma$, i.e., a modified TVS divided by the line flux ($\overline{S}-1$), where $\overline{S}$ is the weighted mean, as defined by \citet{SL09}. Those are presented for the N\,{\sc iv}\,4058, N\,{\sc iii}\,4637, He\,{\sc ii}\,4686, He\,{\sc ii}\,4861, He\,{\sc ii}\,5411, He\,{\sc i}\,5876, He\,{\sc ii}\,6562 and N\,{\sc v}\,6677 lines.

The value of $\Sigma$ quantifies the level of variability, such that a spectrum that reaches a value of $n$ varies with an amplitude $n$ times higher than the variability measured in the continuum (which is assumed to be pure noise). Fig.\,\ref{fig:lpv} shows that all the spectral lines are significantly variable (i.e., $\Sigma (99\%) > 1$), except for the line N{\sc iv}\,4058. \rep{As said above, since this line is closer to the base of the wind than other lines, it is very likely that it is less affected by variations due to CIRs.} Also, if there is any LPV in that line, it would be lost in the noise that tends to be bigger at that wavelength than in the rest of the spectrum, as spectrographs tend to be less efficient at blue wavelengths than at red ones.

For all spectral lines but N\,{\sc iv}\,4058, we can determine a $\sigma$ value, which gives the amplitude of variability relative to the line intensity. By averaging out the $\sigma$ spectrum over the wavelength range corresponding to the emission line, we obtain an average LPV amplitude for the line. Different lines may give different results. We find $\sigma= 5.4 \pm 1.2\%$ (N\,{\sc iii}\,4637), $6.1 \pm 1.2\%$ (He\,{\sc ii}\,4686), $16.5 \pm 3.1\%$ (He\,{\sc ii}\,4861), $8.5 \pm 2.0\%$ (He\,{\sc ii}\,5411), $14.8 \pm 7.5\%$ (He\,{\sc i}\,5876), $7.6 \pm 2.4\%$ (He\,{\sc ii}\,6562) and $7.7 \pm 2.4\%$ (N\,{\sc v}\,6677). The uncertainties are determined by the standard deviation of the $\sigma$ spectrum. We conclude that most lines show comparable $\sigma$-values between 6\% and 8\%, except for He\,{\sc ii}\,4861 and He\,{\sc i}\,5876, that have much higher values. 

The He\,{\sc ii}\,4861 line is more variable because in addition to the CIR-like LPV that can be seen in isolated lines such as He\,{\sc ii}\,5411, the H$\beta$ lines from the two O companions are also moving following the orbit we will describe later in this paper. Of course, the He\,{\sc ii}\,6562 line is itself affected by the H$\alpha$ lines of the O companions, but because this emission line is much stronger than the He\,{\sc ii}\,4861 line, the effect is much smaller. The He\,{\sc i}\,5876 line is potentially also affected by the same line in absorption from the O companions. 

\rep{We can suspect that the LPV have the same origin as the photometric variations with the frequency $\nu_2$=0.615(1)~d$^{-1}$ determined in Section\,\ref{photo}.} Unfortunately, the time sampling of the LPV prevents any robust determination of a period \rep{to compare with}. Indeed, CIR-like LPV are known to be epoch-dependent, and observations that are obtained with a time difference longer than a few weeks cannot be combined, as the configuration of the CIRs at the origin of the LPV has changed completely \citep{SL20}. The typical structure found in the residuals are a few 100 km\,s$^{-1}$ wide, but are wider in the absorption component of P\,Cygni profiles. Indeed, discrete absorption components (DACs) can be observed in P\,Cygni profiles and they vary with the same period as the bumps on top of emission lines \citep{Al16}. The LPV in well-isolated lines, such as He\,{\sc ii}\,4686 and He\,{\sc ii}\,5411, are very similar, while LPV in double lines and multiplex can be interpreted as the same variations repeated as many times as there are contributing lines. 

Finally, the LPVs associated with DACs in the absorption component of the He\,{\sc i}\,5876 line seem to appear in most of other lines as well, indicating that all lines have a component equivalent to a P\,Cygni profile, even when it is filled in by the emission component and is not clearly visible. \rep{This is interesting considering that atmospheric modelling often struggles to reproduce the extended blue absorption profile of strong emission lines where the P Cygni absorption feature seems overestimated \citep[e.g.,][or even in this study]{He01,Sa14,Aa22}. From our observations, we suggest the possibility that it is the emission feature that is underestimated, yet, we cannot at this point offer any physical explanation in support of this statement.}

\section[]{The inner O$+$O eclipsing binary}\label{binary_param}

We use the PHOEBE \citep[PHysics Of Eclipsing BinariEs, version 2.3,][]{prsa05,prsa16, horvat18, conroy20} code to simultaneously model the TESS light and the RV curves. We adopt 1.0 for the bolometric albedos $A$ (coefficients of reprocessing of the emission of a companion by ``reflection'') and for the gravity darkening coefficients $g$. For the limb darkening, we use the square-root law, better suited for hotter stars \citep{Diaz-Cordoves95}. We also account for reflection effects and adopt two reflections. The orbital period was fixed to twice that derived from the TESS light-curve since the eclipses are separated by half the orbital period. As mentioned above, the orbit was adopted to be circular. 

\begin{figure}
\includegraphics[width=8.4cm]{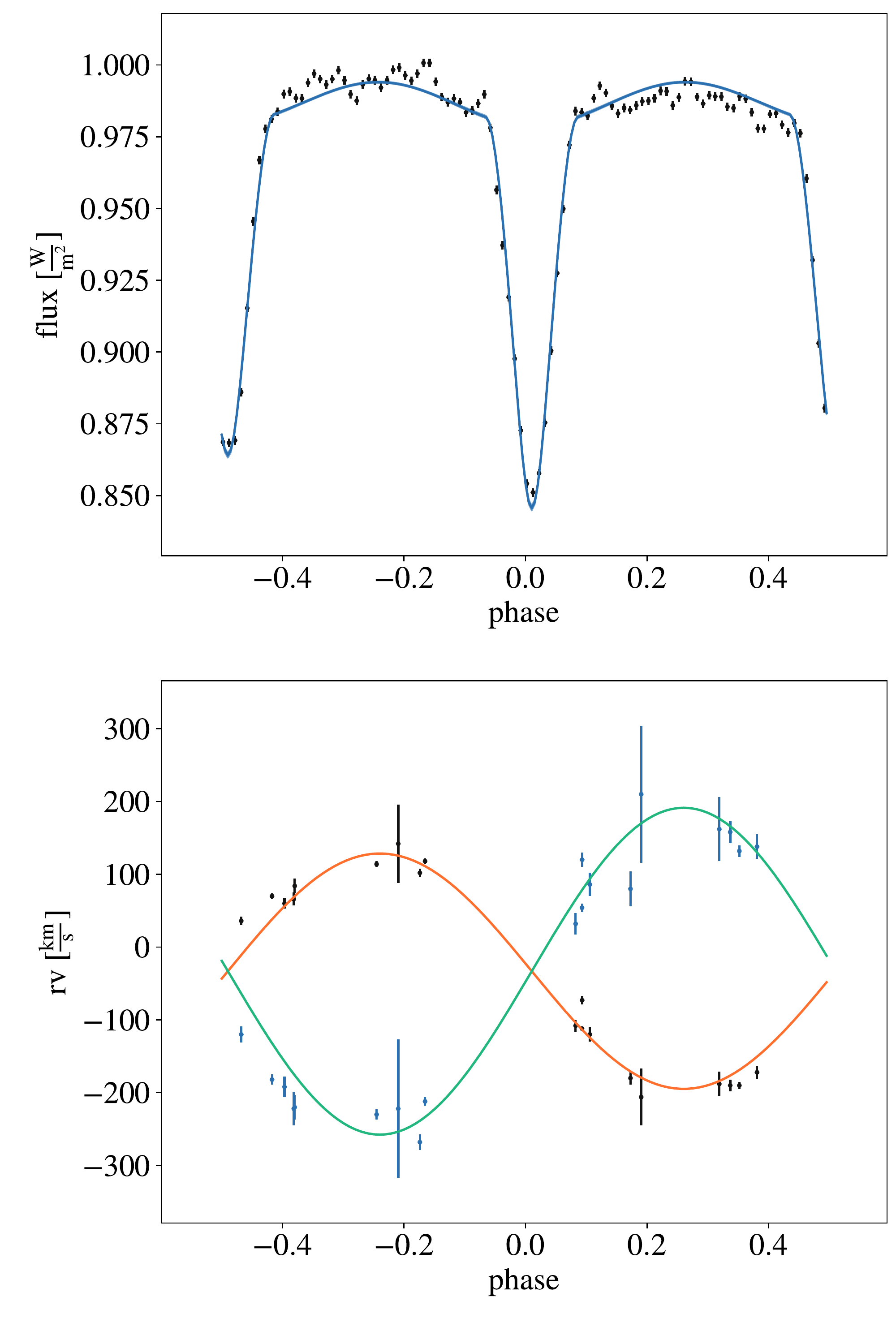}
  \caption{Top: Mean TESS light-curve of WR~63 phase-folded with the 4.03\,d period (black dots) and the fit (blue line) obtained from the orbital fitting using PHOEBE. Bottom: Radial velocity measurements and best-fit RV curves of the O$+$O binary according to the solution listed in Table\,\ref{tab:parameters}. Primary (secondary) is shown in black dots/orange line (blue dots/green line). Phase 0 corresponds to the primary eclipse. \label{fig:phoebe}}
\end{figure}

The problem with WR\,63's light-curve is the high level of variability of the WR wind that confuses the signal from the real eclipses. To prevent the PHOEBE algorithm from getting confused, especially in the interval between the eclipses, we use the median light-curve folded in the orbital period as the input. The uncertainties on the derived parameters are estimated using the Markov Chain Monte Carlo (MCMC) method implemented via emcee \citep{fo13}. Two models have been run to derive the properties of the system: first, a model without including a third light source that would come from the WR star, and a second with it. 

\begin{table}
\caption{\label{tab:orbit} Orbital parameters for the 4.03\,d inner binary, obtained from the orbital analysis using PHOEBE. Error bars correspond to $1\sigma$.}
\centering
\begin{tabular}{lcc}
\hline\hline
Parameter &  \multicolumn{2}{c}{Value} \\
\hline
$P_{\rm orb}$ & \multicolumn{2}{c}{$4.0275 \pm 0.0008$}\\
HJD$_0$ & \multicolumn{2}{c}{$2459000.48169 \pm 0.02083$}\\
$e$& \multicolumn{2}{c}{0.0 (fixed)}\\
$\omega$ [$^\circ$] & \multicolumn{2}{c}{270.0 (fixed)}\\
$q (M_2/M_1)$& \multicolumn{2}{c}{$0.72_{-0.02}^{+0.08}$}\\
$K_1$ [km s$^{-1}$]& \multicolumn{2}{c}{$161.62 \pm 6.50$}\\
$K_2$ [km s$^{-1}$]& \multicolumn{2}{c}{$224.47 \pm 11.20$}\\
$\gamma$ [km s$^{-1}$]& \multicolumn{2}{c}{$-33.13 \pm 3.98$}\\
$M~\sin^3\,i$ [$M_{\odot}$] & $ 13.59 \pm 1.36 $ & $9.92 \pm 0.84 $\\ [1pt] 
$a~\sin\,i$ [$R_{\odot}$] & $ 12.87 \pm 0.47 $ & $17.63 \pm 0.78$\\ [1pt]
\hline\hline
\end{tabular}
\end{table}

Without the third star, the solution yields a system seen under an inclination of $73 \fdg 8^{+0.1}_{-0.1}$. The O-star primary in the inner system has a radius equal to $7.7_{-0.2}^{+0.1}\,R_{\odot}$ and a mass of $15.8_{-0.1}^{+0.1}\,M_{\odot}$. The O-star secondary has a radius of $6.6_{-0.1}^{+0.1}\,R_{\odot}$ and a mass of $11.4_{-0.1}^{+0.1}\,M_{\odot}$. The surface gravities are equal to $3.87^{+0.01}_{-0.01}$ [cgs] for the primary and $3.86^{+0.01}_{-0.01}$ [cgs]  for the secondary. We derive luminosities of $\log(L/L_{\odot}) = 4.45 \pm 0.07$ for the primary, and $\log(L/L_{\odot}) = 4.32 \pm 0.07$ for the secondary. From the observed composite spectra, we also computed the projected rotational velocities of the two stars to be equal to $90_{-10}^{+10}$~km~s$^{-1}$ and $80_{-10}^{+10}$~km~s$^{-1}$, respectively. Using the inclination of the system (and assuming that the stars have a rotational axis perpendicular to the orbital plane), we find the two stars to be in synchronous co-rotation with the orbit. 

\begin{table*}
\caption{\label{tab:parameters} Stellar parameters of the three components in WR~63. Error bars correspond to $1\sigma$.}
\centering
\begin{tabular}{l|cc|cc|c}
\hline\hline
Parameter &  Primary$^{a}$ & Secondary$^{a}$ & Primary$^{a}$ & Secondary$^{a}$ &  WR star$^{b}$ \\
          &  \multicolumn{2}{c}{no third light source} & \multicolumn{2}{c}{with third light source} & \\
\hline
Spectral type & O9~V & OB~V & O9~V & OB~V & WN7\\
$i$ [$^\circ$] & \multicolumn{2}{c}{$73.8^{+0.1}_{-0.1}$} & \multicolumn{2}{c}{$82.9^{+1.0}_{-0.6}$} & $-$\\
$T_{\rm eff}$ [kK] & $31.0 \pm 1.5$ & $26.5 \pm 3.0$ & $31.0 \pm 1.5$ & $30.0_{-4.0}^{+0.8}$ & $42.7 \pm 1.5^{c}$ \\ [1pt] 
$\log (L/L_{\odot})$ & $4.45 \pm 0.07$ & $4.32 \pm 0.07$  & $4.66 \pm 0.08$ & $4.55_{-0.23}^{+0.05}$ & $5.33 \pm 0.14$ \\ [1pt] 
$\log g$ [cgs] & $3.87^{+0.01}_{-0.01} $  & $3.86^{+0.01}_{-0.01}  $ & $3.85^{+0.01}_{-0.01}$  & $3.82^{+0.02}_{-0.02}$ & $-$  \\ [1pt] 
$R$ [$R_{\odot}$] & $7.7_{-0.2}^{+0.1} $ & $6.6_{-0.1}^{+0.1}$ & $7.4_{-0.1}^{+0.1}$ & $7.0_{-0.1}^{+0.1}$ & $8.5_{-1.5}^{+1.7\,c}$ \\ [1pt] 
$M$ [$M_{\odot}$] & $15.8_{-0.1}^{+0.1}$ & $11.4_{-0.1}^{+0.1}$ & $14.3_{-0.1}^{+0.1}$ & $10.3_{-0.1}^{+0.1}$ & $-$\\ [1pt] 
$v~\sin\,i$ [km s$^{-1}$] & $ 90 \pm 10 $ & $80 \pm 10 $ & $ 90 \pm 10 $ & $80 \pm 10 $ & $-$\\ [1pt] 
\hline\hline
\end{tabular}
\rep{ 
\newline
\footnotesize{$^{a}$obtained from PHOEBE modelling, 
$^{b}$obtained from CMFGEN modelling, $^{c}$corresponding to $T_{*} = 44.1 \pm 1.5$~kK and $R_{*} = 7.9_{-1.5}^{+1.7}~R_{\odot}$.}}
\end{table*}

In the second model (i.e., including a third light source), the WR star is expected to be responsible for $64\pm 2$\% of the total flux of the system. As a consequence of the inclusion of the third light source the system is now seen under an inclination of $82 \fdg 9^{+1.0}_{-0.6}$. The radius of the primary is equal to $7.4_{-0.1}^{+0.1}\,R_{\odot}$, and that of the secondary  $7.0_{-0.1}^{+0.1}\,R_{\odot}$. These values are thus similar in the two models. The masses are equal to $14.3_{-0.1}^{+0.1}\,M_{\odot}$ for the O-star primary and $10.3_{-0.1}^{+0.1}\,M_{\odot}$ for the O-star secondary. The inferred surface gravities are $3.85^{+0.01}_{-0.01}$ [cgs] and $3.82^{+0.02}_{-0.02}$ [cgs], for the primary and secondary, respectively. Assuming an effective temperature of $31.0 \pm 1.5$~K and $30.0_{-4.0}^{+0.8}$~K for the primary and secondary, respectively, the computed luminosities are $\log(L/L_{\odot}) = 4.66 \pm 0.08$ and $4.55_{-0.23}^{+0.05}$. The two models yield very similar parameters, even though, when the third light source is not considered, the masses of the two inner components are higher. While both models are acceptable, the sections between the eclipses are better modelled if we include the third light. A summary of the different parameters is given in Table~\ref{tab:parameters}.

\section{Evolutionary status}\label{Evolution}

The analysis of the light-curve provides us with good estimates of the stellar parameters of the two O stars in WR~63. We adopted synthetic spectra for the O stars, using effective temperatures, surface gravities, and projected rotational velocities, as given in Table~\ref{tab:parameters}. The models have been successively convolved with a rotation profile to mimic their rotational velocities ($v~\sin~i = 90$ and 80~km~s$^{-1}$ for the O primary and O secondary, respectively), and a radial-tangential profile to reproduce their macroturbulent velocity ($v_{\rm mac} = 30$~km~s$^{-1}$ for both O stars). \rep{The wind parameters were adopted as follows: the mass-loss rates from \citet{vink01} for Galactic metallicity, the terminal wind velocities are estimated to be 2.6 times the effective escape velocity from the photosphere ($v_{\rm esc}$, \citealt{lamers95}), and the exponent of the velocity law ($\beta$) to be 1 \citep{repolust04}. The models were computed by including the following elements: H, He, C, N, O, Ne, Mg, Al, Si, S, Ar, Ca, Fe, and Ni and the abundances were adopted to be solar \citep{grevesse10}. }

To derive the fundamental properties of the WR star, we used the atmosphere code CMFGEN \citep{hillier98}. CMFGEN is a non-LTE radiative transfer code, developed to model the atmospheres of stars with stellar winds. The luminosity, the radius, the mass-loss rate, the terminal wind speed, and the CNO surface abundances were considered as free parameters. 

\begin{figure*}
\includegraphics[width=17cm]{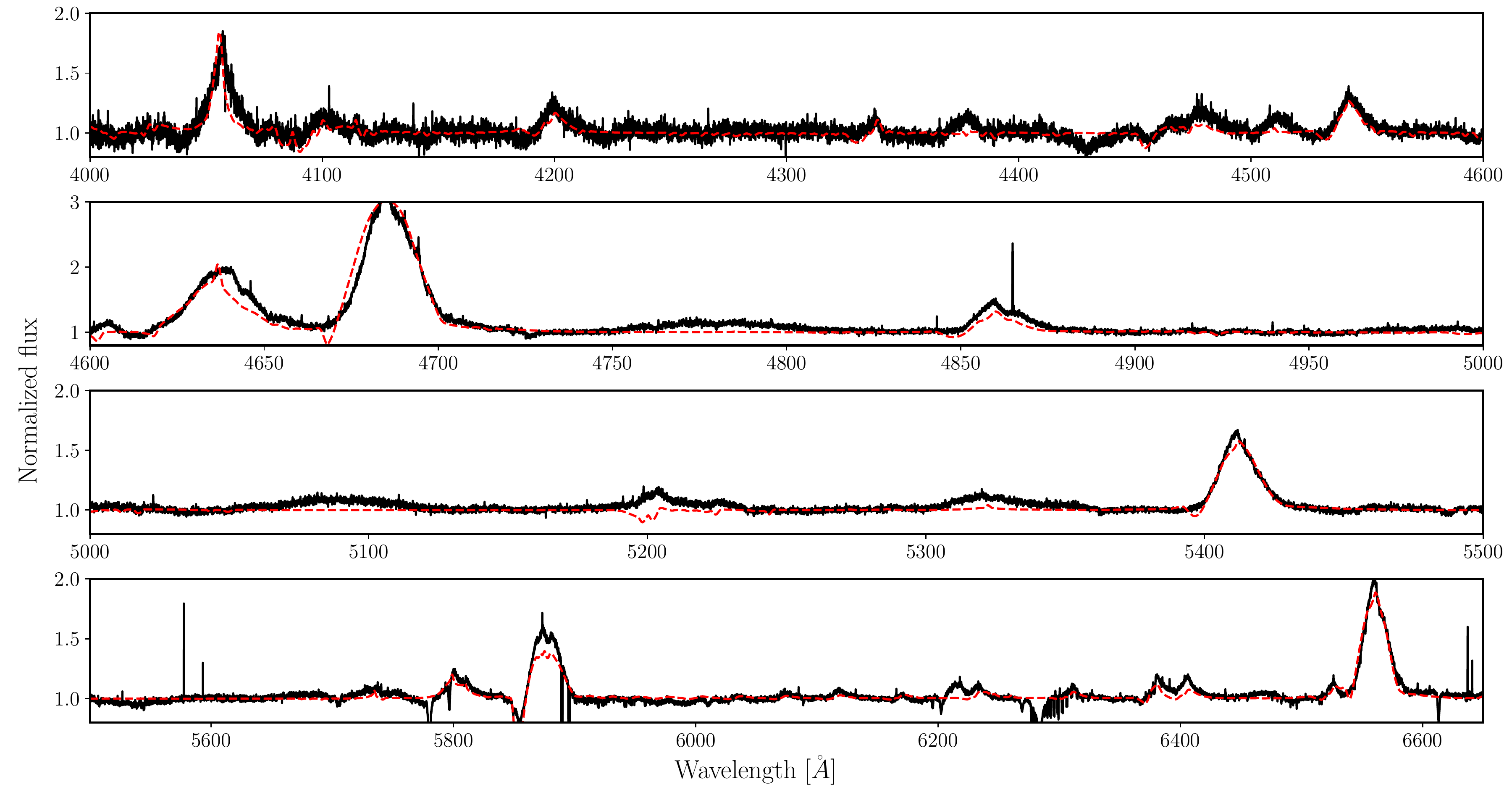}
  \caption{Best-fit CMFGEN model (red) of WR~63 compared to the spectrum acquired on May 1st 2012. The synthetic spectra of the three stars have been combined with dilution factors of 0.65, 0.21, and 0.14 for the WR, the O primary and the OB secondary, respectively. }  \label{fig:CMFGEN}
\end{figure*}

We fitted the spectral energy distribution (SED) of the whole system combining the three SEDs of individual objects. We obtained an extinction of $A_V = 5.31 \pm 0.08$ using the reddening law of \citet{fitzpatrick07}. Given that {\it{Gaia}} does not provide a distance for WR~63, we used the luminosities of the two O stars, the bolometric corrections (computed from \citet{martins05} using their effective temperatures), and the dilution factors of $f_{\rm WR}=0.65 \pm 0.05$, $f_{{\rm{O}}_1}=0.21 \pm 0.03$, and $f_{{\rm{O}}_2}=0.14 \pm 0.02$ for the WR, the O primary and the O secondary, respectively, to compute a distance of $3.4 \pm 0.5$~kpc. Adopting that distance, we compute a stellar luminosity of $\log(L/L_{\odot}) \sim 5.33 \pm 0.14$ for the WR star. All parameters are given in Table~\ref{tab:parameters}.

The modelling was carried out directly by comparing the combined synthetic spectra to the composite spectra. Given the heterogeneity of our observations (different resolutions, wavelength ranges, etc), we decided not to apply spectral disentangling at this point, but to wait for higher-resolution, higher $S/N$ spectra that will be better sampled over the orbital cycle of the O$+$O (inner) binary. We present in Fig.~\ref{fig:CMFGEN} the comparison between one of our observed spectra and the combined synthetic models. Given that the spectra contain three objects, we use the dilution factors listed above to scale the synthetic spectra to the different profiles in the composite spectra.

\rep{Assuming the usual $\beta$ law ($v = v_{\infty}~(1 - R/r)^{\beta}$) with $\beta =1$, we } derive for the WR star an effective temperature of $42.7 \pm 1.5$~kK and a radius of $12.2_{-1.5}^{+1.7}~R_{\odot}$ at $\tau_{\rm ross} = 2/3$. We use a mass-loss rate of $\dot{M} = 10^{-4.80}~M_{\odot}$\,yr$^{-1}$, and a terminal wind speed of $v_{\infty} = 1350$~km~s$^{-1}$. \rep{The synthetic models have been computed by including wind clumping ($f_{\rm cl} = 0.1$). We also compute surface abundances of carbon, nitrogen and oxygen to be $\epsilon_C = 7.78$ (mass fraction: $X(C) = 1.77 \times 10^{-4}$), $\epsilon_N = 9.58$ (mass fraction: $X(N) = 1.31 \times 10^{-2}$), and $\epsilon_O = 8.18$ (mass fraction: $X(O) = 5.90 \times 10^{-4}$), respectively, for the WR star.} We stress that no indication for the presence of hydrogen is detected in the spectrum of this WR star, as mentioned by \citet{hamann06}. This suggests that it has already reached the classical WR phase, and has left the main sequence. \rep{The parameters we derived for the WR star are in agreement with those provided by \citet{hamann06,hamann19} within the error bars. These authors did however not account for the multiplicity of the system, which led to a larger luminosity for the WR star. Given the absence of UV spectra, the terminal wind speed was derived by \citet{hamann06,hamann19} through a careful inspection of the lines, without however specifying which lines were considered. For our analysis, we mainly focus on the P-Cygni profiles visible in the optical, mainly He\,{\sc i}\,5876 (see Fig.~\ref{fig:CMFGEN}). The error bars on the terminal wind speed ($\sim 250$~km~s$^{-1}$) are therefore larger than when one derives it from the UV P-Cygni profiles. }

\begin{figure*}
\includegraphics[width=18cm]{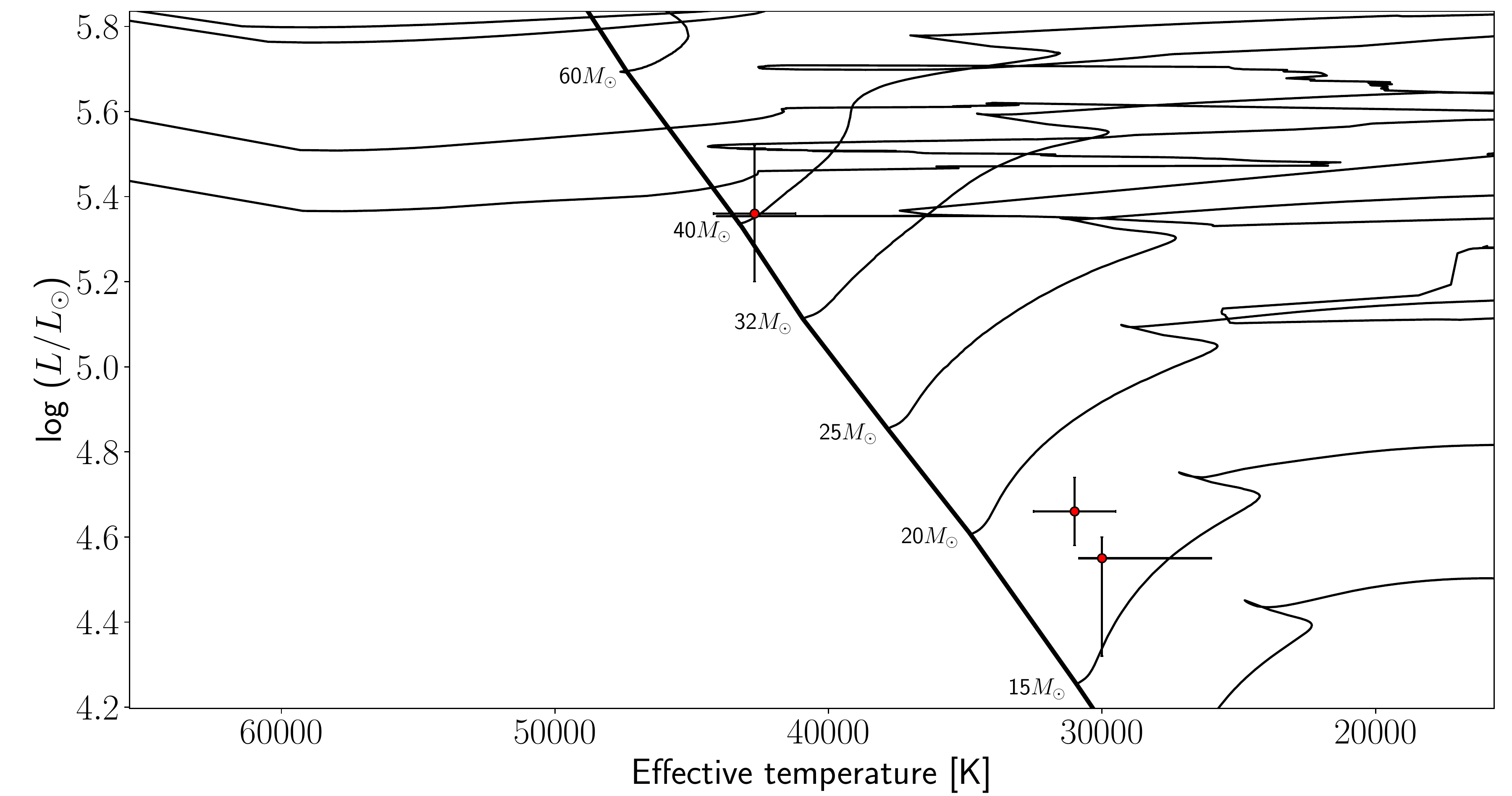}
  \caption{Hertzsprung-Russell diagram. The tracks are from \citet{ekstrom12}, computed with an initial rotational velocity of 300~km~s$^{-1}$. The locations of the three objects are based on the PHOEBE model including a third light source. \label{fig:HRD}}
\end{figure*}

\rep{The evolutionary status of the three stars is interpreted using the single-star tracks from \citet{ekstrom12}, with an initial rotational velocity of $\sim 300$~km~s$^{-1}$, as they predict the stellar evolution until the helium core-burning phase, unlike tracks from Bonn \citep{brott11}}. We use the derived parameters for the O stars using the model with the third light source (see Table\,\ref{tab:parameters}), and those of the WR star as input to estimate the ages of the three stars. We display in Fig.~\ref{fig:HRD} a Hertzsprung-Russell diagram, showing the positions of the three stars. We interpolated the stellar parameters through the Geneva tracks to estimate the predicted ages of the three stars. Within the error bars, it appears that the three stars are coeval (i.e., they have the same age). That indicates that the three objects have evolved as single stars. According to the Geneva tracks, the WR star is expected to have an initial mass of $35 \pm 5~M_{\odot}$ while the primary O star and the secondary O star are expected to have initial masses of $18 \pm 2$ and $16 \pm 2$~$M_{\odot}$, respectively. The Geneva tracks indicate an age of $5.9 \pm 1.4$~Myrs for the WR star, $6.0 \pm 1.1$~Myrs for the O primary and $8.2 \pm 2.5$~Myrs for the O secondary. One should note that the effective temperature of the secondary O star has a large uncertainty, and more spectroscopic data are needed to confirm our results. 

Our spectroscopic data covers 3317 days, divided into five periods. Over that time span, we were not able to detect any motions of the WR star or the combined inner O$+$O binary system. Assuming that the three stars have evolved as single stars in a hierarchical system, the WR star should have gone through a red supergiant or luminous blue variable phase. During this phase, the radius of the star is assumed to reach hundreds of solar radii. The outer orbit (that hosting the WR star) must have a semi-major axis as large as that value, which implies the need for a long orbital period. That could indeed explain why, from our data, we are unable to detect the motion of the WR star. This is also true if the outer orbit is highly eccentric. Therefore, more observations are needed to constrain the real nature of the system, but, if WR~63 is really a hierarchical system, it could serve as a Rosetta stone for testing the single-star evolutionary tracks.

\section{Summary and conclusions}\label{Conclusions}

Using high-resolution spectra and high-cadence photometry, we have identified three components in the composite spectra of WR~63: an O9~V primary and an O9.5~V secondary that are paired in a 4.03\,d orbit, and a WR star classified as WN7. Our analysis does not allow us to show that the WR star is gravitationally bound to the O$+$O system. Even though a spurious alignment between the WR star and the O$+$O system is possible, the likelihood of having a hierarchical triple system seems more plausible. 

We have shown that the two O stars form a detached binary system, and that there is no sign that these components have already interacted with each other. Assuming that the WR star is bound to the inner system, the age of the system can be inferred by the WR star. Using the Geneva tracks \citep{ekstrom12}, the mean age of WR~63 is estimated to be of about $5.9 \pm 1.4$~Myrs. Combining the estimated age with the masses and luminosities of the O stars, we showed that the three objects are coeval. 

Longer-term observations are therefore required to shed more light on the exact structure of this complex system, specifically elucidating the remaining hierarchical configuration of the WR star with respect to the O$+$O system. Knowing the orbital and physical properties of such a system could serve as corner stone to study the Kozai-Lidov mechanism that occurs in hierarchical triple systems.

\section*{Acknowledgments}
\rep{The authors would like to thank the anonymous referee whose constructive comments helped improving this publication.}
Based on observations obtained at the international Gemini Observatory, a program of NSF’s NOIRLab, which is managed by the Association of Universities for Research in Astronomy (AURA) under a cooperative agreement with the National Science Foundation, on behalf of the Gemini Observatory partnership: the National Science Foundation (United States), National Research Council (Canada), Agencia Nacional de Investigaci\'{o}n y Desarrollo (Chile), Ministerio de Ciencia, Tecnolog\'{i}a e Innovaci\'{o}n (Argentina), Minist\'{e}rio da Ci\^{e}ncia, Tecnologia, Inova\c{c}\~{o}es e Comunica\c{c}\~{o}es (Brazil), and Korea Astronomy and Space Science Institute (Republic of Korea).

L.M.\ thanks the European Space Agency (ESA) and the Belgian Federal Science Policy Office (BELSPO) for their support in the framework of the PRODEX Programme.

E.G.\ is also greatly indebted to the same institutions for PRODEX programmes linked to the
XMM-Newton and {\it{Gaia}} missions. He is also thankful to the Belgian F.R.S.-FNRS for various supports.
N.S.L.\ acknowledges financial support from the Natural Sciences and Engineering Research Council (NSERC) of Canada.
K.D. thanks the European Research Council under European Union’s Horizon 2020 research programme (grant agreement No. 772225).

The TESS data presented in this paper were obtained from the Mikulski Archive for Space Telescopes (MAST) at the Space Telescope Science Institute (STScI), which is operated by the Association of Universities for Research in Astronomy, Inc., under NASA contract NAS5-26555. Support to MAST for these data is provided by the NASA Office of Space Science via grant NAG5-7584 and by other grants and contracts. Funding for the TESS mission is provided by the NASA Explorer Program. 

This research has made use of the SIMBAD database, operated at CDS, Strasbourg, France and of NASA's Astrophysics Data System Bibliographic Services. We are grateful to the staff of the ESO La Silla and Paranal Observatories for their technical support. This paper is based in part on spectroscopic observations made with the Southern African Large Telescope (SALT) under program 2021-1-SCI-013 (PI: Manick). We are grateful to our SALT colleagues for maintaining the telescope facilities and conducting the observations. 

\section{Data availability}
Most spectroscopic observations underlying this article will have past the proprietary period by the time this publication will be published, and raw data can be downloaded from the ESO, SALT and Gemini archives.
The reduced spectra and the light-curves will be shared on reasonable request to the corresponding author.


\end{document}